\documentclass[12pt]{iopart}
\usepackage{float}
\usepackage{graphicx}
\usepackage{amsfonts}

\expandafter\let\csname equation*\endcsname\relax
\expandafter\let\csname endequation*\endcsname\relax
\usepackage{amsmath} 

\usepackage[T1]{fontenc}
\usepackage{bm}
\usepackage{amssymb}
\usepackage{leftidx}
\usepackage{upgreek}
\usepackage{xcolor}
\usepackage[outdir=./]{epstopdf}

\usepackage{iopams}  

\begin{document}

\title[Enhancement of electron magnetic susceptibility due to many-body interactions in monolayer MoSe$_2$]{Enhancement of electron magnetic susceptibility due to many-body interactions in monolayer MoSe$_2$}
\author{K.~Oreszczuk$^1$, A.~Rodek$^1$, M.~Goryca$^1$, T.~Kazimierczuk$^1$, M.~Raczy\'nski$^1$,
J. Howarth$^2$, T.~Taniguchi$^3$, K.~Watanabe$^4$, M.~Potemski$^{1,5,6}$, and
P.~Kossacki$^1$} 

\address{$^1$ Institute of Experimental Physics, Faculty of Physics, University of Warsaw, ul. Pasteura 5, 02-093 Warsaw, Poland}
\address{$^2$ National Graphene Institute, University of Manchester, Booth St E, Manchester, M13 9PL, United Kingdom}
\address{$^3$ International Center for Materials Nanoarchitectonics, National Institute for Materials Science, 1-1 Namiki, Tsukuba 305-0044, Japan}
\address{$^4$ Research Center for Functional Materials, National Institute for Materials Science, 1-1 Namiki, Tsukuba 305-0044, Japan}
\address{$^5$ Laboratoire National des Champs Magn\'etiques Intenses, UPR 3228, CNRS, EMFL, Univ; Grenoble Alpes, 38042 Grenoble, France}
\address{$^6$ CENTERA Labs, Institute of High Pressure Physics, PAS, 01 - 142 Warszawa, Poland}

\ead{Kacper.Oreszczuk@fuw.edu.pl}
\vspace{10pt}
\begin{indented}
\item[]March 2023
\end{indented}

\begin{abstract}
Employing the original, all-optical method, we quantify the magnetic susceptibility of a two-dimensional electron gas (2DEG) confined in the MoSe$_2$ monolayer in the range of low and moderate carrier densities. The impact of electron-electron interactions on the 2DEG magnetic susceptibility is found to be particularly strong in the limit of, studied in detail, low carrier densities. Following the existing models, we derive the value of  $g_0 = 2.5 \pm 0.4$ for the bare (in  the absence of the interaction effects) $g$-factor of the ground state electronic band in the MoSe$_2$ monolayer. The derived value of this parameter is discussed in the context of estimations from other experimental approaches. Surprisingly, the conclusions drawn differ from theoretical ab-initio studies.
\end{abstract}

%
\vspace{2pc}
\noindent{\it Keywords}: TMD, MoSe$_2$, 2DEG, magnetic susceptibility, $g$-factor,
%
\submitto{\TDM}
%
%
%

\section{Introduction}

Collective properties of the 2D electron gas (2DEG) and the 2D hole gas (2DHG) have been studied extensively for a wide variety of systems. The ease of the electrical control of carrier concentration makes two-dimensional systems an excellent playground for exploring carrier-carrier interactions, facilitating many interesting phenomena, such as spin textures \cite{sondhi1993Phys.Rev.B} or quantum Hall ferromagnetism \cite{girvin2000Phys.Today}. The Coloumb repulsion enhances the pseudospin susceptibility \cite{girvin2000Phys.Today, jungwirth2000Phys.Rev.B, ando1974J.Phys.Soc.Jpn., dassarma2009Phys.Rev.B, attaccalite2002Phys.Rev.Lett.}, possibly leading to the spontaneous creation of broken symmetry phases \cite{wojs2002Phys.Rev.B, giuliani1985Phys.Rev.B, scrace2015Nat.Nanotechnol., li2020Phys.Rev.Lett.a}.

Monolayers of transition metal dichalcogenides (TMDs) draw a lot of attention as semiconducting materials with robust optical properties, strong Coulomb interaction \cite{you2015Nat.Phys., he2014Phys.Rev.Lett., ross2013Nat.Commun.} and an optically accessible valley degree of freedom \cite{mak2010Phys.Rev.Lett., splendiani2010NanoLett., zeng2012Nat.Nanotechnol., mak2012Nat.Nanotechnol.}. TMDs are an example of a system with near-ideal two-dimensional quantization, where the magnetooptical measurements reveal significant collective effects \cite{li2020Phys.Rev.Lett., larentis2018Phys.Rev.B}. Relatively large carrier masses and reduced dielectric screening \cite{mak2016Nat.Photon., wang2018Rev.Mod.Phys.} promise strong electron-electron interactions even at large carrier concentrations. 

Collective behavior of the carriers in an ideal 2DEG can be modeled, for example, employing the Quantum Monte Carlo (QMC) calculations \cite{attaccalite2002Phys.Rev.Lett.}. Theoretical methods predict a strong enhancement of the carrier gas polarizability in low carrier density regimes. The experimentally obtained values of the carrier gas polarizability can be compared with theoretical predictions to estimate the effective single-particle band $g$-factor\cite{larentis2018Phys.Rev.B}. 

Phenomena related to the enhancement of the spin susceptibility or other collective properties of the 2D carrier gas in monolayer TMDs
can be experimentally observed through the Landau level filling processes in both Zeeman-split valleys \cite{li2020Phys.Rev.Lett., movva2017Phys.Rev.Lett., larentis2018Phys.Rev.B, pisoni2018Phys.Rev.Lett., wang2017Nat.Nanotechnol., smolenski2019Phys.Rev.Lett., liu2020Phys.Rev.Lett., gustafsson2018Nat.Mater.}. Such experiments usually yield large effective $g$-factor ($g^*$) values for 2DEG in MoS$_2$\cite{lin2019NanoLett., larentis2018Phys.Rev.B}, 2DHG in WSe$_2$\cite{gustafsson2018Nat.Mater., li2020Phys.Rev.Lett., movva2017Phys.Rev.Lett.} and 2DEG in MoSe$_2$\cite{larentis2018Phys.Rev.B}. The $g$-factor values usually reach between $10$ and $20$ for carrier densities close to $4\cdot10^{12}\,\text{cm}^{-2}$. This approach, however, is viable only in the regime of relatively high carrier densities 
(usually above $3\cdot10^{12}\,\text{cm}^{-2}$).
The low and moderate carrier density regimes, where the collective effects are expected to be most pronounced, must be investigated by different approaches. In that view, the experimental techniques that are viable in low carrier densities are of high importance.   

Here we present a novel experimental approach to study the polarizability of 2DEG in the MoSe$_2$ monolayer by means of all-optical electron gas polarization measurements. We employ two complementary approaches to measure the effective electron $g$-factors in low and
moderate 2DEG densities. Our experimental methods are not constrained by the Landau quantization and can be employed both in high and low-to-moderate 2DEG density regimes. Moreover, our approaches are more versatile by permitting the measurement of materials with significant inhomogeneity or Coloumb disorder, which otherwise would prohibit the application of methods relying on the signatures of the Landau levels.

\section{Sample and experimental Setup}
\begin {figure}[] \begin {center} 
\includegraphics{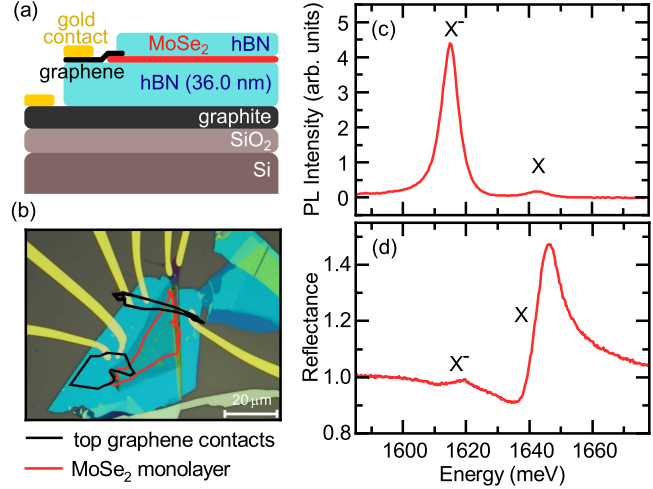}
\caption{ (a) Scheme of the MoSe$_2$ monolayer with graphene contacts on the top and graphite gate at the bottom. (b) Photograph of the sample. The positions of the MoSe$_2$ monolayer and graphene top contacts are marked with contours. The graphite back gate is visible as a darker area in the middle of the hBN flake. (c) PL spectrum of the sample under the excitation with $600$\,nm, $10\,\mu$W femtosecond pulsed laser. (d) Reflectivity spectrum of the sample. Spectra in (c-d) were acquired at the gate voltage of $0$\,V in the temperature of $6$\,K. Neutral exciton (X) and negatively charged exciton (X$^-$) resonances can be resolved.}
\label {sample} \end {center}\end {figure}

The electrically gated MoSe$_2$ sample, schematically shown in Fig \ref{sample}(a), was prepared with the means of mechanical exfoliation using the dry transfer method. The MoSe$_2$ monolayer was exfoliated on top of the graphite back gate and $36$\,nm hBN spacer. The thickness of the hBN spacer was measured with the atomic force microscope. Next, the second graphene flake was then deposited at the edge of the MoSe$_2$ monolayer to provide electric contact. Finally, gold contacts were deposited on the graphite back gate and on the graphene contact. The photograph of the sample is shown in Fig \ref{sample}(b).
The photoluminescence and reflectance spectra of the MoSe$_2$ monolayer are presented in Fig. \ref{sample}(c-d). The reflectance spectra were divided by the reference spectrum measured on the gold contact and then normalized.
Neutral exciton (X) and negatively charged exciton (X$^-$) resonances can be resolved in the photoluminescence and reflectance spectra at the zero gate bias voltage. 

Most magnetooptical experiments were carried out in a~cryostat equipped with
a~superconducting coil providing magnetic fields up to $10$\,T. In this experimental setup, samples were placed in a helium gas atmosphere at
the temperature of $6$\,K or $11$\,K or in the superfluid helium bath at $1.6$\,K. The temperature was calibrated with a thin film resistive sensor placed on the sample holder. Unless noted otherwise, the data were acquired at $6$\,K. The optical signal was focused by a single aspheric lens (NA~$=0.68$) placed inside the cryostat.

High magnetic field (up to $30$\,T) measurements were performed with the use of a resistive magnet at the temperature of $4.2$\,K. The optical signal was focused by a microscope objective (NA~$=0.25$) placed inside the cryostat.

In both experimental configurations, the sample was mounted on the x-y-z piezoelectric stage.

\section{Results}

\begin {figure}[] \begin {center} 
\includegraphics{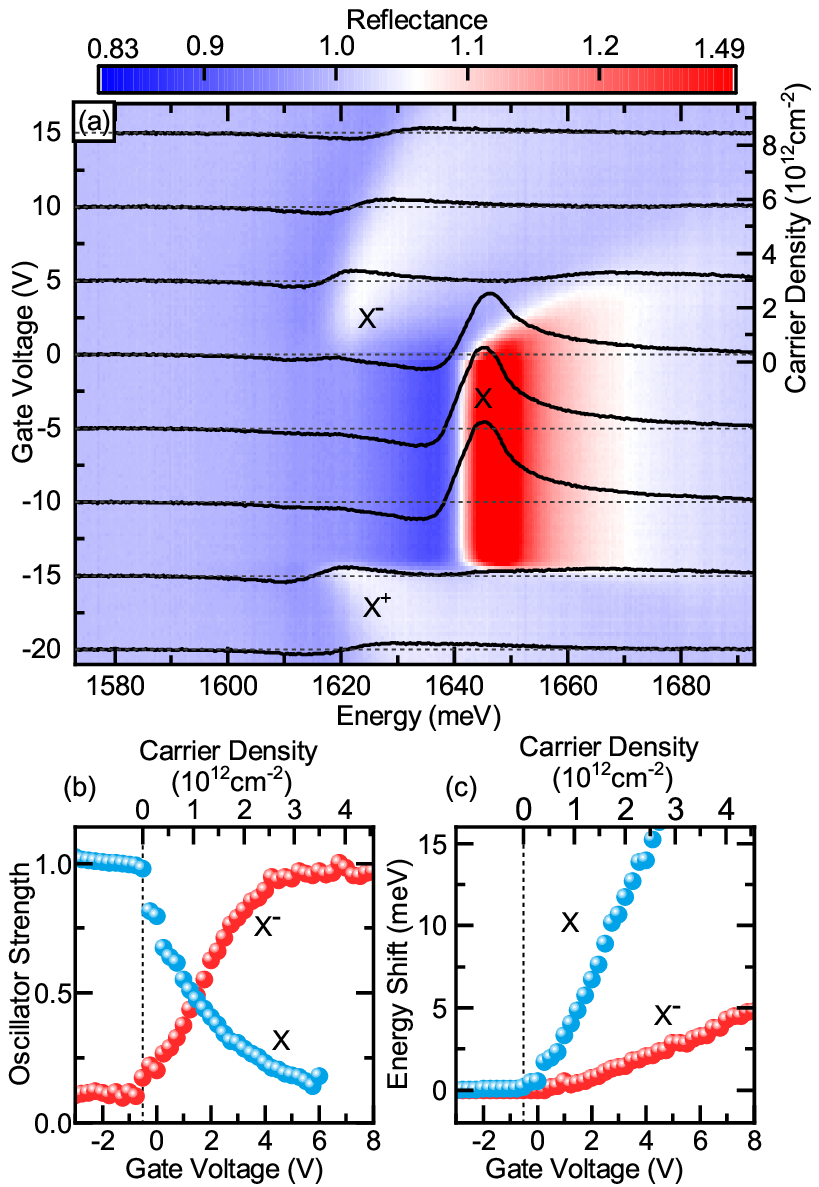}
\caption{ (a) Reflectance spectrum of the MoSe$_2$ monolayer measured at different gate voltages in the temperature of $6$\,K. Spectra are composed of neutral exciton (X), negatively charged exciton (X$^-$) and positively charged exciton (X$^+$) resonances. (b) Oscillator strengths of the neutral and charged exciton peaks normalized to their maximal value. (c) Energy shift of the X and X$^-$ peaks relative to their positions in the neutral regime. Vertical dashed line in (b) and (c) marks the voltage of $-0.5$\,V, where the Fermi level crosses the bottom of the conduction band.}
\label {map} \end {center}\end {figure}

The map of the reflectance spectrum of the sample plotted against the gate voltage is presented in Fig. \ref{map}(a). Three carrier concentration regimes can be distinguished. The p-doping regime emerges at strong negative gate voltages, while the moderate negative gate voltages result in the mid-bandgap Fermi level, yielding a neutral regime. Positive gate voltage induces n-doping of the MoSe$_2$ monolayer.

The exact value of the n-doping transition voltage is important for the precise determination of the electron densities. Figs. \ref{map}(b-c) shows the amplitudes and positions of the excitonic peaks as a function of the gate voltage. Qualitative change is visible at the voltage of $-0.5$\,V, indicating a transition between neutral and n-doped regimes. The negative value of the transition voltage indicates that the studied sample is in the weak 2D electron gas regime at zero bias voltage. The 2D hole gas regime can be distinguished at gate voltages below $-14$\,V. 

In this work, we focus on the attractive polaron/negatively charged exciton (X$^-$) peak in the 2DEG\ regime. The carrier concentration is approximated with a flat capacitor formula with thickness equal to $d=36.0$\,nm. The estimation of the relative permeability of hexagonal boron nitride $\epsilon_{\perp}\approx3.5$  is based on previous works 
\cite{kim2012ACSNano,laturia2018npj2DMater.Appl.}. The exact value of the threshold voltage between the 2DEG regime and the neutral regime was determined from the changes in the oscillator strength and resonance energy of the X$^-$ 
peak (Fig \ref{map}(b-c)). 

Using two complementary methods, we investigate the effective susceptibility of the electron gas at different carrier densities. We neglect the Landau quantization in both experimental approaches and treat the electron bands as continuous. The Landau level separation equals $1.4$\,meV at the magnetic field of $10$\,T. Consequently, the resulting electron density in a single Landau level is equal to $0.24\,\cdot10^{12}\text{cm}^{-2}$ , with both values scaling linearly with the magnetic field. The Landau level splitting is significantly smaller than the Zeeman splittings analyzed in these experiments. In particular, the Landau level separation is
also smaller than the amplitude of the electric potential fluctuations within the monolayer. This prohibits Landau-level signatures from emerging in the experimental data. 

\begin {figure}[] \begin {center} 
\includegraphics{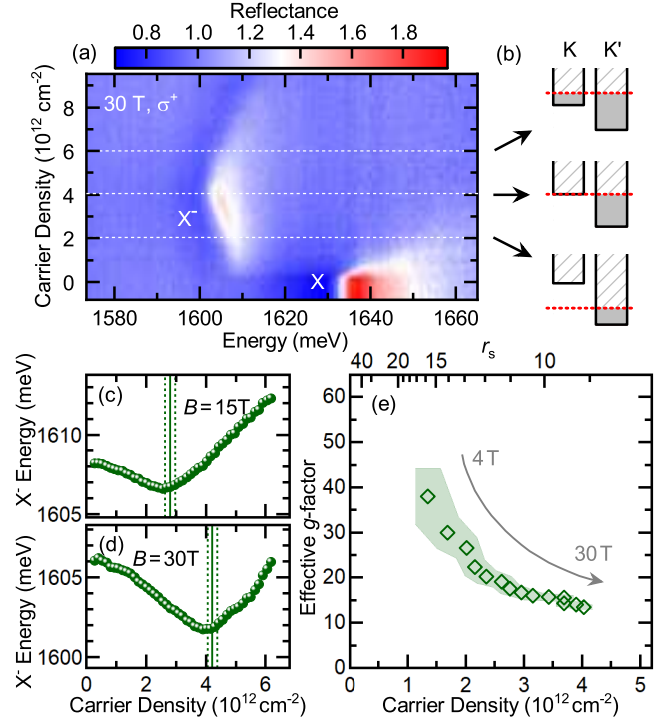}
\caption{  (a) Reflectance spectrum of  the MoSe$_2$ monolayer at different carrier densities in the 2DEG regime in the magnetic field of $30$\,T. (b) Schematic presentation of the valley filling at three different carrier densities marked in panel (a). (c-d) Dependence of the X$^-$ resonance energy on the carrier density. Vertical lines mark the threshold carrier density for filling the second Zeeman-split valley. (e) Effective electron $g$-factors are calculated from threshold carrier densities measured at different magnetic fields.  }
\label {gfactors_grenoble} \end {center}\end {figure}

 In the primary approach, we apply the magnetic field up to $30\,$T and probe the X$^-$ resonance at different 2DEG densities at $\sigma^+$ polarization of detection (Fig. \ref{gfactors_grenoble}(a)). At sufficiently high magnetic fields, two different regimes can be distinguished. At low carrier densities, 2DEG is fully confined in the energetically favorable $K^-$ valley, while above a certain density threshold, both valleys are partially filled (Fig. \ref{gfactors_grenoble}(b)). The two regimes can be differentiated by observing the rate of the change of the energy of the X$^-$ resonance upon increasing 2DEG density. Below the threshold density, the resonance energy decreases with increasing density due to the attractive interactions with the electron gas. Above the threshold density, when the $K^+$ valley starts being filled, k-space filling effects induce a rapid positive shift of the X$^-$ resonance energy (Fig. \ref{gfactors_grenoble}(c-d)). We measure the threshold 2DEG density in different magnetic fields. Then we calculate the Zeeman splitting necessary to contain 2DEG in the single valley and, consequently, determine the effective electron $g$-factor. Here, we define the effective $g$-factor $g^*$ so that the Zeeman Splitting $E_Z=g^*\mu_B B$, in line with the definition used in 
\cite{li2020Phys.Rev.Lett.,gustafsson2018Nat.Mater.} (note, that the definition used in 
\cite{larentis2018Phys.Rev.B,movva2017Phys.Rev.Lett.,lin2019NanoLett.} differs by a factor of 2). The effective mass of the electron in monolayer MoSe$_2$ is assumed equal to   $m_e=0.84$ (Goryca et al.~\cite{goryca2019Nat.Commun.}), from which follows the single-valley density of states $\rho= \frac{m_0 m^*}{2\pi \hbar ^2}$. The values of the effective electron $g$-factor at different carrier densities are presented in Fig. \ref{gfactors_grenoble}(e).

To complete the determination of the electron $g$-factor in a broader range of carrier densities, we employ a supplementary approach. We focus on the circular polarization resolved oscillator strength of the X$^-$ resonance in the MoSe$_2$ monolayer in the 2D electron gas regime. Fig. \ref{map}(a) shows the reflectance spectrum of the sample at different gate voltages.    
   
\begin {figure} \begin {center} 
\includegraphics{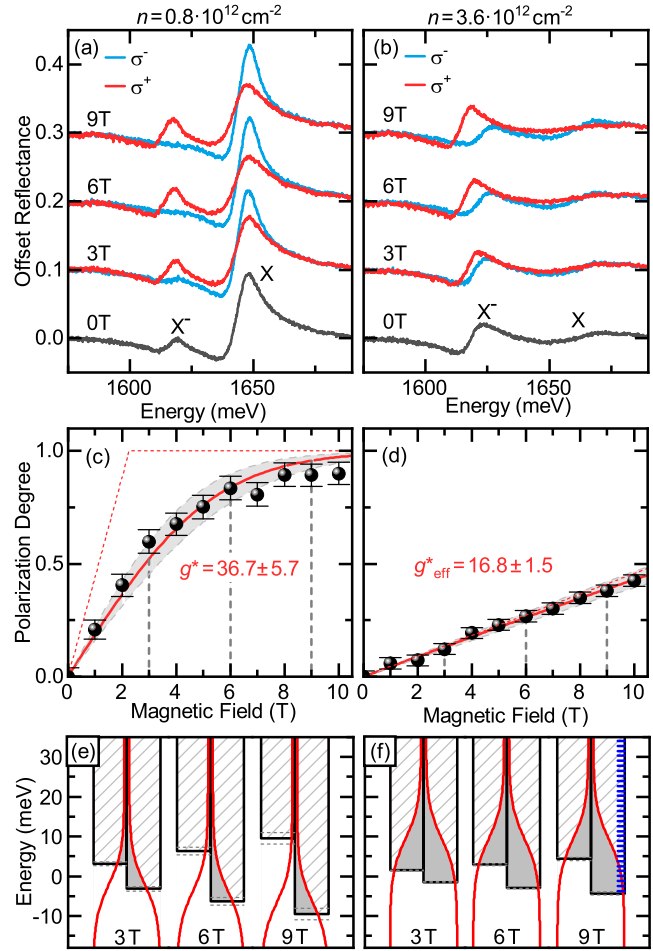}
\caption{ (a-b) Reflectance spectra of the MoSe$_2$ monolayer measured with the circular polarizations of detection at different magnetic fields in the temperature of $6$\,K. Voltages applied to the gated structure are equal to $1$\,V (a) and  $6$\,V (b), resulting in 2DEG density equal to $0.84\cdot10^{12}\,\text{cm}^{-2}$ and $3.6\cdot10^{12}\,\text{cm}^{-2}$, respectively.  (c-d) Magnetic field dependencies of circular polarization degree of the X$^-$ peak at corresponding 2DEG densities. Red solid lines: fits to the experimental data with the effective $g$-factor value as a fitting parameter. Red dashed lines: polarization degrees expected under conditions of no disorder and very low temperature. (e-f) Visualizations of the Zeeman-split valleys at different magnetic fields at corresponding 2DEG densities. Red lines visualize the Fermi-Dirac distribution at the temperature of $6$\,K. Carrier density at each valley is proportional to the area of the shape filled with gray color.  Horizontal dashed lines correspond to the unceratinity of the $g^*$-induced Zeeman splitting. Blue dashes on panel (f) visualize the energy separation of the Landau Levels at the magnetic field of $9$\,T. }
\label {fits} \end {center}\end {figure}

The reflectance of the sample is probed with the circular polarization of detection at opposite magnetic fields, corresponding to the $\sigma^+$ and $\sigma^-$ polarizations (Fig \ref{fits}(a-b)).
The reflectance measurements are performed at different magnetic fields and gate voltages. We analyze the oscillator strengths of transitions related to   X$^-$ formation in two circular polarizations (oscillator strengths are denoted as $I^+$ and $I^-$ in  $\sigma^+$  and $\sigma^-$  polarizations, respectively). Figs. \ref{fits}(c-d) show the magnetic field dependence of the circular polarization degree ($P=(I^+-I^-)/(I^++I^-)$) of the X$^-$ resonance at two different carrier concentrations. 

We describe the magnetic field evolution of the observed circular polarization degree of the X$^-$ peak with a simple model based on the following assumptions: 
\begin{itemize}
\item[(i)] 
The total carrier density is obtained from the flat capacitor formula, which means that the Zeeman shifts are negligible with respect to electrostatic potential. 
\item[(ii)]
The effective mass of the electron in monolayer MoSe$_2$ is $m_e=0.84$ (Goryca et al.~\cite{goryca2019Nat.Commun.}), from which follows the single-valley density of states $\rho= \frac{m_0 m^*}{2\pi \hbar ^2}$.  
\item[(iii)]
The electron gas follows the Fermi-Dirac distribution with the effective temperature $T_{\text{eff}}$.
\item[(iv)]
The polarization degree $P$ of the 2DEG is equal to the polarization degree of the X$^-$ resonance observed in the reflectance spectrum.\end{itemize}

In assumption (iii), the effective temperature parameter $T_{\text{eff}}$ describes both the distribution related to the nonzero temperature of the electron gas as well as the approximation of the in-plane Coulomb fluctuations. The latter effect is dominant in this work due to the low temperatures. One should note that the electric potential fluctuations in the conductance band may be comparable but not necessarily equal to the inhomogeneous broadening of the excitonic peaks. The effective temperature parameter $T_{\text{eff}}$ is derived from the sample temperature $T$ and conductance band disorder parameter $T_{\text{dis}}$ as a root sum of squares: $T_{\text{eff}}^2 = T^2 + T_{\text{dis}}^2$. The estimated value of the disorder parameter is equal  $T_{\text{dis}} = (50 \pm15)$\,K. The estimation was performed in a way to ensure the quality of fit in Fig. \ref{fits}(e) and to match the values of the effective $g$-factor obtained with the high magnetic field approach. Thus, the results -- which rely strongly on the value of the $T_{\text{dis}}$ parameter -- must be treated as an extrapolation of the high magnetic field approach. As such, the uncertainties of both are strongly correlated. This is relevant mainly in the low carrier density regime (below $2\cdot10^{12}\text{cm}^{-2}$), and has no impact on the results obtained in carrier densities above $4\cdot10^{12}\text{cm}^{-2}$. 
See Supplementary Information, Figs. 1-2 for more information on the impact of the $T_{\text{dis}}$ parameter on the fit quality and resulting values of the effective $g$-factor.  

Assumption (iv) may require additional justification, as the oscillator strength of the X$^-$ resonance is directly proportional to the 2DEG density only in the part of the gate voltages range (Fig. \ref{map}(b)). However, previous works on the 2DEG in quantum wells \cite{kossacki1999Phys.Rev.B} indicate that the observed sublinearity of the oscillator strength of the X$^-$ peak in high carrier density can be described in terms of the reduction factor $\xi(n)$ common for both polarizations of detection and dependent on the total carrier density $n$, rather than in terms of separate reduction factors dependent on the carrier densities $n_{K^{+(-)}} $\ attributed to each valley. In this description, the oscillator strength at $\sigma^+(\sigma^-)$ polarization of detection is described by formula $I^{+(-)} = n_{K^{+(-)}} \cdot \xi(n)$, leading to the equivalence between optically observed polarization degree and the distribution of the carriers in Zeeman-split valleys: $P=(I^+-I^-)/(I^++I^-) = (n_{K^+}-n_{K^-})/(n_{K^+}+n_{K^-})$. 

Such assumptions are further supported by the observation that the total oscillator strength of the X$^-$ resonance remains constant for 2DEG polarizations induced by different magnetic fields (Figs. \ref{fits}(a-b)). In particular, we find the oscillator strength at $B=0\,$T to be equal to half the oscillator strength at the saturation magnetic field. Reduction factor -- if common for both circular polarizations -- does not perturb the observed polarization degree of the X$^-$ resonance, assuring the validity of the assumption (iv).

 Carrier densities in  $K^+$ and $K^-$ valleys can be described with the integral of the Fermi-Dirac distribution over the single-band density of states $\rho$: 
\begin{equation}
n_{K^{+(-)}}=\int_{\pm\frac{E_{Z}}{2}}^{\infty } \frac{\rho}{e^{\frac{E-\mu}{k T_{\text{eff} }}}+1} \, dE = \rho k T_\text{eff} \log\left( e^{\frac{\pm\frac{E_{Z}}{2}}{k T_{\text{eff} }}}+e^{\frac{\mu }{k T_{\text{eff} }}} \right)\mp\rho  \frac{E_{Z}}{2} 
\label{gfactors_npm} \end{equation}
There is exactly one solution for the Fermi level $\mu$ and Zeeman splitting $E_Z$ that satisfies the observed magnetic polarization degree $P = (n_{K^+}-n_{K^-})/(n_{K^+}+n_{K^-})$  and the total carrier density $n=n_{K^{+}}+n_{K^{-}}$ obtained from the flat capacitor formula. Similarly, the effective electron $g$-factor $g^*$ (from which follows the Zeeman splitting $E_Z$) can be treated as a fit parameter describing the 2DEG polarization degrees observed in the experiment.

Figs. \ref{fits}(c-d) show the polarization degree of the X$^-$ reflectance peak plotted against the magnetic field at two different 2DEG densities. The effective electron $g$-factor was calculated for each carrier density to reproduce the observed polarization degree dependency on the magnetic field. Figs. \ref{fits}(e-f) show the visualizations of the Zeeman Splitting and carrier distribution between K$^+$\ and K$^-$ valleys for different 2DEG densities and magnetic fields. Red solid lines on Figs. \ref{fits}(e-f) represent the expected polarization degree evolution according to the best fit of the effective $g$-factor to the experimental data. 

\begin {figure}[] \begin {center} 
\includegraphics{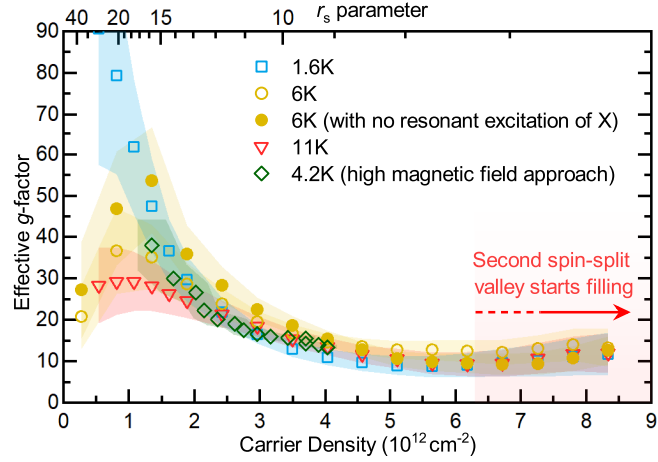}
\caption{ Effective $g$-factor value as a function of the electron gas density measured with different approaches and in different temperatures. }
\label {gfactors} \end {center}\end {figure}

The effective $g$-factor was calculated for several gate voltages at three different temperatures. The results are presented in Fig. \ref{gfactors} with blue, yellow, and red data points. Regardless of the temperature, the effective $g$-factor exhibits a value of more than $30$ at concentrations below $2\cdot10^{12}\,\text{cm}^{-2}$ and declines pronouncedly reaching values close to  $10$ at the 2DEG densities above $5\cdot10^{12}\,\text{cm}^{-2}$. The value of the effective $g$-factor seemingly rebounds in carrier densities above $7\cdot10^{12}\,\text{cm}^{-2}$. However, this probably is an artifact related to the filling of the second spin-split valley. The spin-orbit splitting in the conduction band of the MoSe$_2$ monolayer is equal to $\Delta_C=21\,$meV \cite{kosmider2013Phys.Rev.B}, which corresponds to the carrier density of $7\cdot10^{12}\,\text{cm}^{-2}$ distributed between the two valleys.

Recent work \cite{smolenski2022Phys.Rev.Lett.} suggests that resonant excitation of neutral exciton peak can induce significant depolarization of the 2DEG, even under the weak excitation induced by white light. To verify the influence of  that effect in our experimental conditions, we performed an additional measurement with the white light spectral range limited to the nearest proximity of the X$^-$ peak, that is, to 
$1560-1633$\,meV. The spectral range was fine-tuned during the measurement to follow the energy shifts of the X and X$^-$ peaks, present when the magnetic field and gate voltage were not equal to zero.
The results of this experiment are presented in Fig. \ref{gfactors} with filled yellow circles. We do observe an increase in polarizability after filtering out the resonant excitation, however, the effect is incremental and does not fundamentally alter our experimental results.

\begin {figure} \begin {center} 
\includegraphics{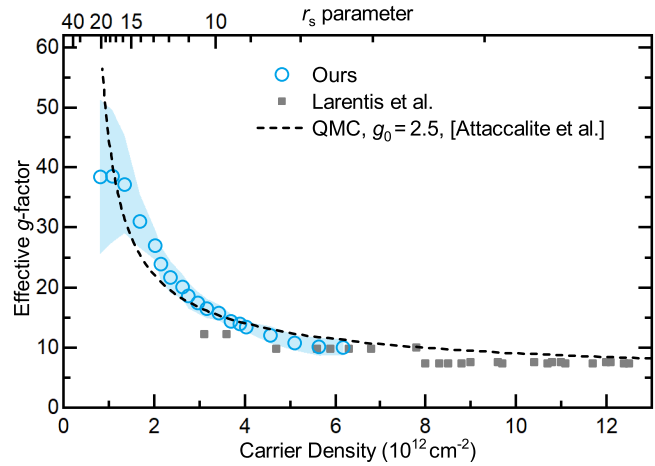}
\caption{ Circles: Values of the effective electron $g$-factor in monolayer MoSe$_2$ averaged over different approaches and different temperatures presented in Fig.~\ref{gfactors}. Dashed line: fit of the QMC calculations \cite{attaccalite2002Phys.Rev.Lett.} to the experimental data, resulting in the value of the electron $g$-factor in the absence of interaction effects $g_0 = 2.5 \pm 0.2$. Squares: Values obtained in Landau Level studies by Larentis et al. \cite{larentis2018Phys.Rev.B} (resulting in $g_0=2.2$).}
\label {gfactorsfinal} \end {center}\end {figure}

The results of all our experiments performed in different temperatures and with different approaches have been averaged and presented in Fig. \ref{gfactorsfinal}. Wherever the $g$-factor value at given carrier density was missing in some of the averaged datasets, linear interpolation between nearest data points was used. 
We fit the QMC calculations \cite{attaccalite2002Phys.Rev.Lett.} and obtain the value of the lowest conduction band $g$-factor in the absence of the interaction effects equal to  $g_0=2.5\pm0.4$. The interparticle distance parameter $r_s$ was calculated with the effective dielectric constant of the hBN environment equal to $\kappa=\sqrt{\epsilon_\perp \epsilon_\parallel}$, assuming $\epsilon_\parallel=6.9$ \cite{laturia2018npj2DMater.Appl.}. 

Our results are in agreement with Landau Level studies in monolayer MoSe$_2$ \cite{larentis2018Phys.Rev.B} ($g_0=2.2$). Our approaches, however, allow for probing deeper into the low carrier density regime, where the collective effects are most pronounced.
The value of the conduction band $g$-factor obtained in our work is also in line with the experimental estimations made for monolayer MoSe$_2$ in the neutral regime \cite{koperski20182DMater.} ($g_0\approx2.2$). Here, however, one should note that the $g$-factor describing magnetic susceptibility differs from the electron factor used (together with the hole one) to calculate the Zeeman splitting of the optical transition. In particular, the Kohn theorem states that optical transitions are not sensitive to carrier-carrier interactions when the carriers have the same effective mass \cite{perez2007Phys.Rev.Lett., boukari2006Phys.Rev.B}. 

Our results, supported by experimental approaches from other works, create a solid image of the issue of the conductance band $g$-factor in monolayer MoSe$_2$. These results, however, are in disagreement with the results of the Ab Initio calculations, which predict up to twofold greater values ($g_0=3.6$ \cite{ wozniak2020Phys.Rev.B},
 $g_0=3.8$ \cite{forste2020NatCommun},
 $g_0=4.1$ \cite{xuan2020Phys.Rev.Res.},
 $g_0=5.5$ \cite{deilmann2020Phys.Rev.Lett.}). The observed discrepancy suggests room for improvement in the theoretical models or assumptions of the Ab Initio calculations. This may relate either to the predictions on the band $g$-factor value or to the validity of the assumptions on the 2DEG interactions. 

Some degree of uncertainty may arrive from the determination of the $r_s$ parameter, which is derived from the Bohr radius of the electron, which consequently depends on the assumptions taken on the dielectric environment of the electrons. In particular, the dielectric constant of hBN exhibits strong boundary effects resulting in significant thickness dependence of its dielectric constant \cite{laturia2018npj2DMater.Appl.}. Although significant for the quality of fit, uncertainties regarding the $r_s$ parameter scale (see Fig. \ref{gfactorsfinal}) remain unlikely to influence the resulting $g_0$ value significantly.


\section{Conclusions}
We determined the effective electron magnetic susceptibility in low and
moderate 2DEG densities in the monolayer MoSe$_2$. Our measurements were performed with two different approaches and at different temperatures between $1.6$\,K and $11$\,K. 
All obtained results are quantitatively consistent. The exploration deep into the low carrier density regime reveals strong collective effects. The shape of the susceptibility variation with electron density agrees with the QMC calculations \cite{attaccalite2002Phys.Rev.Lett.} and leads to the value of the lower electron band $g$-factor in the absence of the interaction effects: $g_0=2.5\pm0.4$. This value, while coherent with other experimental approaches \cite{larentis2018Phys.Rev.B,koperski20182DMater.}, contests the results of to-date Ab-Initio calculations  \cite{ wozniak2020Phys.Rev.B, deilmann2020Phys.Rev.Lett., forste2020NatCommun, xuan2020Phys.Rev.Res.},  opening perspectives for improvement in assumptions and theoretical models.
 
\ack
This work was supported by National Science
Centre, Poland under projects 2021/41/N/ST3/04240 and 2020/39/B/ST3/03251. The research leading to these results has also received funding from the Norwegian Financial Mechanism 2014-2021 within project No. 2020/37/K/ST3/03656 and from the Polish National Agency for Academic Exchange within Polish Returns program under Grant No. PPN/PPO/2020/1/00030. We acknowledge the support of the LNCMI-CNRS, member of the European Magnetic Field Laboratory (EMFL)  and of the CNRS via IRP "2DM" project.
M.P acknowledges the support from EU Graphene Flagship and from FNP-Poland (IRA - MAB/2018/9 grant, SG 0P program of the EU).\section*{References}

\bibliographystyle{iopart-num}
\bibliography{elemose2}

\end{document}